\documentclass[english]{article}
\usepackage[T1]{fontenc}
\usepackage[latin1]{inputenc}
\usepackage{geometry}
\usepackage{amsmath}
\usepackage{graphicx}
\usepackage{setspace}
\usepackage{bm}
\usepackage{amssymb}
\usepackage{epsf}

\makeatletter
 \newcommand{\lyxaddress}[1]{
   \par {\raggedright #1
   \vspace{1.4em}
   \noindent\par}
 }

\usepackage{babel}
\makeatother
\begin{document}

\title{Long-range entanglement in the Dirac vacuum}

\author{J. Silman and B. Reznik}

\maketitle

\lyxaddress{\begin{center}\emph{School of Physics and Astronomy, Raymond and
Beverly Sackler Faculty of Exact Sciences, Department of Physics and
Astronomy, Tel-Aviv University, Tel-Aviv 69978, Israel }\end{center}}

\begin{abstract}
\textbf{Recently, there have been a number of works investigating
the entanglement properties of distinct noncomplementary parts of
discrete and continuous Bosonic systems in ground and thermal states.
The Fermionic case, however, has yet to be expressly addressed. In
this paper we investigate the entanglement between a pair of far-apart
regions of the 3+1 dimensional massless Dirac vacuum via a previously
introduced distillation protocol [B. Reznik, \emph{et
al.}, Phys. Rev. A 71, 042104 (2005)]. We show that entanglement
persists over arbitrary distances, and that as a function of $L/R$,
where $L$ is the distance between the regions and $R$ is their typical
scale, it decays no faster than $\sim exp\left[-\left(L/R\right)^{2}\right]$.
We discuss the similarities and differences with analogous results
obtained for the massless Klein-Gordon vacuum.}

\end{abstract}
\vspace{1.3em}
Entanglement in spatially extended many body systems and quantum field
theories is the focus of increasing attention. Part of this is directed
at understanding the entanglement properties of noncomplementary parts
of a system, such as far apart regions of vacuum \cite{Clifton,Benni,VBI,Verch,Many-region}
and thermal states \cite{Braun,Massar}, or widely separated segments
in ground states of the Bose-Hubbard Hamiltonian \cite{Lewenstein},
chains of trapped ions \cite{Ion traps} and harmonic oscillators
\cite{Vedral}. In this regard relativistic vacua are especially interesting
as they provide us with an oppurtunity to study physical systems with
a well defined notion of locality.

In this paper we investigate the entanglement between arbitrarily
distant regions of the free massless Dirac vacuum. For Bosonic systems,
the expansion of the vacuum in terms of two-mode squeezed states of
oscillators residing in the two complementary spacetime wedges $x>0$
and $x<0$, used in the derivation of the Unruh effect, explicitly
shows that the vacuum is entangled \cite{Unruh,Summers}. This result
is a special case of a general modewise decomposition theorem pertaining
to a certain class of Bosonic Gaussian states \cite{Modewise,Giedke}.
An analogous theorem exists for Fermionic Gaussian states \cite{Modewise 2}.
(Indeed, the Unruh effect holds also in the Fermionic vacuum \cite{Unruh2}.)
The state of a pair of noncomplementary parts of a system, however,
in general is mixed, so that a modewise decomposition is impossible
\cite{note0}. Working directly with the system's degrees of freedom,
especially when of a great or inifinite number, proves difficult then.
A most effective and relatively simple way to tackle this problem
is the use of entanglement distillation protocols. Even though such
protocols have proved most convenient in the study of the entanglement
between abitrarily distant regions of the Bosonic vacuum \cite{VBI,Verch,Many-region},
the Fermionic case has thus far not been expressly addressed. Using
a previously introduced distillation protocol \cite{Benni,VBI}, we
explicitly show that results analogous to those obtained for the Bosonic
vacuum are true of the Fermionic vacuum as well, namely, that entanglement
persists between arbitrarily far-apart regions and that as a function
of the ratio of the separation between the regions $L$ and their
typical scale $R$, the entanglement decays no faster than $\sim exp\left[-\left(L/R\right)^{2}\right]$.
\vspace{1.3em}

The concepts of entanglement and locality are nontrivial in Fermionic
systems. We therefore begin by explaining them briefly and contrast
with the Bosonic case. Suppose we have a system of Bosonic modes.
As $\left[a_{i},\, a_{j}^{\dagger}\right]=\delta_{ij}$ and $\left[a_{i},\, a_{j}\right]=0$,
the Hilbert space is a direct product of the Hilbert spaces of each
of the modes. Hence, it is meaningful to consider the entanglement
between different sets of modes, with the partition unequivocally
defining locality. For Fermions $\left\{ a_{i},\, a_{j}^{\dagger}\right\} =\delta_{ij}$
and $\left\{ a_{i},\, a_{j}\right\} =0$. The Hilbert space therefore
lacks an analogous direct product structure. If the assignation of
sets of modes to different parties is to have any meaning at all (that
is, if we do not want to give up locality), we must restrict the set
of observables in such a way that acting on an arbitrary composite
state of any two distinct sets of modes, say $A$ and $B$, with an
observable comprised solely of modes in $A$, does not change the
expectation values of any observable comprised solely of modes in
$B$ (and vice-versa), nor increase the entanglement between the sets.
Now, let $\hat{O}_{A}$ and $\hat{O}_{B}$ be arbitrary sums of products
of even number of modes in $A$ and $B$, respectively, then $\left[\hat{O}_{A},\,\hat{O}_{B}\right]=0$.
It is not hard to see that this characteristic of Fermionic modes,
together with the anti-commutation algebra of the modes, restricts
the set of observables to those that can be constructed out of products
of an even number of modes \cite{Bravyi}. For pure states entanglement
between two sets of modes is then defined as usual, i.e. a pure composite
state of two sets of modes, $\phi$, is entangled iff there exist
observables $\hat{O}_{A}$ and $\hat{O}_{B}$ such that $\left\langle \hat{O}_{A}\hat{O}_{B}\right\rangle _{\phi}\neq\left\langle \hat{O}_{A}\right\rangle _{\phi}\left\langle \hat{O}_{B}\right\rangle _{\phi}$.
Of course this is not true of mixed states. Indeed, except in $2\times2$
and $2\times3$ dimensions \cite{Peres,Horodecki}, no necessary and
sufficient criterion to establish mixed state entanglement is known,
regardless of the statistics.

Moving on to relativistic quantum field theory (QFT), the requirement
of Lorentz covariance and that the energy spectrum be bounded from
below constrains the set of possible algebras of modes to the familiar
commutation/anti-commutation relations for Bosons/Fermions. As an
example consider the Dirac field \begin{equation}
\left\{ \psi_{i}\left(\vec{x},\, t\right),\,\psi_{j}^{\dagger}\left(\vec{y},\, t\right)\right\} =\delta_{ij}\delta\left(\vec{x}-\vec{y}\right),\qquad\left\{ \psi_{i}\left(\vec{x},\, t\right),\,\psi_{j}\left(\vec{y},\, t\right)\right\} =0.\label{anticommut}\end{equation}
The subscripts denote the spinorial indices, which together with the
position $\vec{x}$ label the modes. If in addition we want the theory
to be causal, we must require that observables be respresented by
bilinear expressions in the fields.

It is important to note that there is a difference in what is meant
by {}``local'' in quantum information theory (QIT) and QFT settings.
As explained above, in QIT it is the different parties that define
locality. However, in QFT it is causality which defines locality,
i.e. an operator acting at two or more spacelike related coordinates
is nonlocal. In this paper, locality in the QIT sense enters via the
assignation of causally disconnected regions to different parties,
leaving us with much greater latitude in our choice of {}``local''
operations than that afforded by the tight constraints of QFT.
\vspace{1.3em}

The protocol employed in \cite{Benni,VBI} consists of the finite
duration coupling of a pair of initially nonentangled two-level point-like
detectors to the studied field, in its vacuum state, at two different
locations. The duration of the coupling determines the size of the
regions {}``probed'' and is taken to be much smaller than the distance
between the detectors, which therefore remain causally disconnected.
Under these conditions, a final entangled state of the detectors means
that entanglement persists between the regions. A similar but suitably
adjusted protocol is employed here. We therefore make use many of
the results obtained in \cite{VBI}, and forego rederivation.

The Dirac equation is given by \begin{equation}
\left(\vec{\alpha}\cdot\vec{p}+\beta m\right)\psi\left(\vec{x},\, t\right)=i\frac{\partial}{\partial t}\psi\left(\vec{x},\, t\right),\label{Dirac eq.}\end{equation}
where the $\alpha_{i}$ and $\beta$ are any $4\times4$ matrices
satisfying $\left\{ \alpha_{i},\,\alpha_{j}\right\} =2\delta_{ij}\mathbf{1}$,
$\beta^{2}=\mathbf{1}$, and $\left\{ \alpha,\,\beta\right\} =0$.
In the absence of a mass term, in the Weyl representation of the $\alpha_{i}$
and $\beta$, the Dirac equation decouples into a pair of equations,
the Weyl equations \begin{equation}
\vec{\sigma}\cdot\vec{p}\,\psi_{r}\left(\vec{x},\, t\right)=i\frac{\partial}{\partial t}\psi_{r}\left(\vec{x},\, t\right),\qquad-\vec{\sigma}\cdot\vec{p}\,\psi_{l}\left(\vec{x},\, t\right)=i\frac{\partial}{\partial t}\psi_{l}\left(\vec{x},\, t\right),\label{Weyl Eq.}\end{equation}
where the $\sigma_{i}$ are the Pauli matrices. The two-component
fields $\psi_{r}\left(\vec{x},\, t\right)$ and $\psi_{l}\left(\vec{x},\, t\right)$
describe right and left handed particles and anti-particles, that
is quanta of positive and negative helicity, respectively. We can
therefore begin by studying a vacuum of definite handedness, say the
right handed vacuum. In terms of a Fourier expansion \begin{equation}
\psi_{r}\left(\vec{x},\, t\right)=\int\frac{d^{3}p}{\left(2\pi\right)^{3}}\left(a_{\vec{p}}^{r}u_{r}\left(\vec{p}\right)e^{-i\left(pt-\vec{p}\cdot\vec{x}\right)}+\left.b_{\vec{p}}^{r}\right.^{\dagger}v_{r}\left(\vec{p}\right)e^{i\left(pt-\vec{p}\cdot\vec{x}\right)}\right).\label{Dirac field}\end{equation}
$\left.a_{\vec{p}}^{r}\right.^{\dagger}$, $a_{\vec{p}}^{r}$ and
$\left.b_{\vec{p}}^{r}\right.^{\dagger}$, $b_{\vec{p}}^{r}$ are
the creation and annihilation operators for right handed particles
and anti-particles, respectively, satisfying $\left\{ a_{\vec{p}}^{r},\,\left.a_{\vec{p}}^{r}\right.^{\dagger}\right\} =\left\{ b_{\vec{p}}^{r},\,\left.b_{\vec{p}}^{r}\right.^{\dagger}\right\} =\left(2\pi\right)^{3}\delta\left(\vec{p}-\vec{q}\right)$
with all other anti-commutators vanishing, while $u_{r}\left(\vec{p}\right)$
and $v_{r}\left(\vec{p}\right)$ are the corresponding two-component
{}``spinorial'' coefficients and are understood to be normalised
to unity.

Due to the fact that the field is Fermionic, we must couple to field
bilinears to realise the distillation protocol \cite{VBI}. (See earlier
discussion.) Perhaps the most natural choice is the field's charge
density $\hat{N}\left(\psi_{r}^{\dagger}\left(\vec{x}_{i},\, t\right)\psi_{r}\left(\vec{x}_{i},\, t\right)\right)$,
where for convenience we have chosen to normal order ($\hat{N}$).
Setting up a pair of two-level detectors at $\vec{x}_{A}$and $\vec{x}_{B}$,
in the Dirac interaction picture the coupling term is given by \begin{equation}
H_{C}\left(t\right)=\frac{1}{2}\sum_{i=A,\, B}\epsilon_{i}\left(t\right)\left(Cos\left(\Omega_{i}t\right)\sigma_{i}^{x}+Sin\left(\Omega_{i}t\right)\sigma_{i}^{y}\right)\hat{N}\left(\psi_{r}^{\dagger}\left(\vec{x}_{i},\, t\right)\psi_{r}\left(\vec{x}_{i},\, t\right)\right).\label{coupling}\end{equation}
Here $\epsilon_{i}\left(t\right)$ governs the strength and duration
of the coupling, and $\Omega_{i}$ is the energy gap of detector $i$.
The corresponding evolution operator is $U\left(T/2\right)=\hat{T}exp\left[-i\int_{-T/2}^{T/2}dtH_{C}\left(t\right)\right]$,
with $\hat{T}$ and $T$ denoting time-ordering and the duration of
the interaction, respectively. As discussed above, we set $L\gg T$
($L:=\left|\vec{x}_{B}-\vec{x}_{A}\right|$), and take the initial
state of the detectors to be separable.

Once the interaction is over, in the basis $\left\{ \downarrow\downarrow,\,\downarrow\uparrow,\,\uparrow\downarrow,\,\uparrow\uparrow\right\} $,
the partial transpose of the detectors' reduced density matrix {\small is
given by \begin{equation}
\rho_{AB}^{PT}=\left(\begin{array}{cccc}
1-\left\Vert E_{A}\right\Vert ^{2}-\left\Vert E_{B}\right\Vert ^{2} & 0 & 0 & \left\langle E_{A}\mid E_{B}\right\rangle \\
0 & \left\Vert E_{B}\right\Vert ^{2} & -\left\langle 0\mid X_{AB}\right\rangle  & 0\\
0 & -\left\langle X_{AB}\mid0\right\rangle  & \left\Vert E_{A}\right\Vert ^{2} & 0\\
\left\langle E_{B}\mid E_{A}\right\rangle  & 0 & 0 & \left\Vert X_{AB}\right\Vert ^{2}\end{array}\right)+O\left(\epsilon_{i}^{3}\right),\label{DM}\end{equation}
}where \begin{equation}
\left|E_{i}\right\rangle :=\int_{-T/2}^{T/2}dt\epsilon_{i}\left(t\right)e^{i\Omega_{i}t}\hat{N}\left(\psi_{r}^{\dagger}\left(\vec{x}_{i},\, t\right)\psi_{r}\left(\vec{x}_{i},\, t\right)\right)\left|0\right\rangle ,\label{emission}\end{equation}

\begin{equation}
\left|X_{AB}\right\rangle :=\int_{-T/2}^{T/2}dtdt'\epsilon_{A}\left(t\right)\epsilon_{B}\left(t'\right)e^{i\left(\Omega_{A}t+\Omega_{B}t'\right)}\hat{N}\left(\psi_{r}^{\dagger}\left(\vec{x}_{A},\, t\right)\psi_{r}\left(\vec{x}_{A},\, t\right)\right)\hat{N}\left(\psi_{r}^{\dagger}\left(\vec{x}_{B},\, t'\right)\psi_{r}\left(\vec{x}_{B},\, t'\right)\right)\left|0\right\rangle .\label{exchange}\end{equation}
Using the Peres criterion \cite{Peres}, we find that the detectors
are entangled (i.e. that the partial transpose has negative eigenvalues)
if \begin{equation}
\left|\left\langle 0\mid X_{AB}\right\rangle \right|^{2}-\left\Vert E_{A}\right\Vert ^{2}\left\Vert E_{B}\right\Vert ^{2}>0.\label{condition}\end{equation}
Physically speaking, this translates to the requirement that the probability
of exchange of a right handed virtual particle - anti-particle pair
between the detectors be greater than the product of the probabilities
for the on-shell emission of a right handed particle - anti-particle
pair by the same detector.

For temporally symmetric window functions, a somewhat lengthy calculation
shows that the above condition takes on the explicit form (see appendix
for details)\begin{eqnarray}
\left|\int_{0}^{\infty}\frac{d\omega}{L^{2}}\omega^{3}Cos\left(\omega L\right)\tilde{\epsilon}_{A}\left(\Omega_{A}+\omega\right)\tilde{\epsilon}_{B}\left(\Omega_{B}-\omega\right)+6\int_{0}^{\infty}\frac{d\omega_{1}d\omega_{2}}{L^{3}}\left[\frac{1}{L}Sin\left(\omega_{1}L\right)Sin\left(\omega_{2}L\right)\right.\right.\nonumber \\
-\omega_{1}Cos\left(\omega_{1}L\right)Sin\left(\omega_{2}L\right)-\omega_{2}Sin\left(\omega_{1}L\right)Cos\left(\omega_{2}L\right)\bigg{]}\tilde{\epsilon}_{A}\left(\Omega_{A}+\omega_{1}+\omega_{2}\right)\tilde{\epsilon}_{B}\left(\Omega_{B}-\omega_{1}-\omega_{2}\right)\bigg{|}^{2} & >\nonumber \\
\frac{1}{25}\int_{0}^{\infty}d\omega\omega^{5}\tilde{\epsilon}_{A}\left(\Omega_{A}+\omega\right)^{2}\int_{0}^{\infty}d\omega\omega^{5}\tilde{\epsilon}_{B}\left(\Omega_{B}+\omega\right)^{2},\label{3+1 condition}\end{eqnarray}
where $\tilde{\epsilon}_{i}$ is the Fourier transform of $\epsilon_{i}$.

This is to be compared with the condition obtained for the massless
real Klein-Gordon field \cite{Benni} \begin{equation}
\left|\int_{0}^{\infty}\frac{d\omega}{L}Sin\left(\omega L\right)\tilde{\epsilon}_{A}\left(\Omega_{A}+\omega\right)\tilde{\epsilon}_{B}\left(\Omega_{B}-\omega\right)\right|^{2}>\int_{0}^{\infty}d\omega\omega\left|\tilde{\epsilon}_{A}\left(\Omega_{A}+\omega\right)\right|^{2}\int_{0}^{\infty}d\omega\omega\left|\tilde{\epsilon}_{B}\left(\Omega_{B}+\omega\right)\right|^{2}\label{3D Klein}\end{equation}
 The two conditions bear similarity. The analysis performed in \cite{VBI}
shows that Eq. (\ref{3D Klein}) can be satisfied if we choose $\tilde{\epsilon}_{A}\left(\Omega_{A}-\omega\right)$
such that it oscillates as $Sin\left(\omega L\right)$, \emph{that
is faster than any of its Fourier components}, over a finite integration
regime \cite{Super,Berry}. Indeed, such a choice \cite{Benni2} can
render the exchange probability arbitrarily larger than the product
of the emission probabilities. Suppose in our case we take the superoscillatory
transform to oscillate like $Cos\left(\omega L\right)$ over a suitably
chosen integration regime. Then the specially tailored form of the
superoscillatory transform guarantees that the first term on the LHS
of Eq. (\ref{3+1 condition}) is much greater than the RHS. But for
precisely the same reason it is much greater than all other terms
on the LHS, and Eq. (\ref{3+1 condition}) is satisfied. It follows
that entanglement persists between arbitrarily far-apart regions of
the massless Dirac vacuum of quanta of definite handedness, and that
the lower bound obtained in \cite{VBI} holds here as well. That is,
in the limit $L/T\gg1$ the entanglement, quantified by the negativity
$\mathcal{N}$, scales no faster than $\sim exp\left[-\left(L/T\right)^{2}\right]$.
Now as in the duration $T$ the detectors probe a spherical region
of radius $R=T$ (see Fig. 1), we arrive at the aforementioned lower
bound $\sim exp\left[-\left(L/R\right)^{2}\right]$.

\begin{figure}
\includegraphics[scale=0.79]{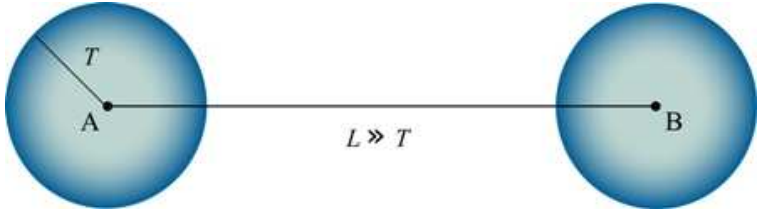}

\caption{Spacetime at the start of the interaction $t=-T/2.$ $A$ and $B$
denote the locations of the detectors. It is only operators within
the two spheres that contribute to the distilled entanglement.}
\end{figure}

Not surpirsingly, for the left handed vacuum the condition for entanglement
is identical. This means that double the amount of entanglement can
be distilled by coupling to the total charge density, that is the
sum of the charge densities of right and left handed quanta.
\vspace{1.3em}

Before we conclude, there are three questions that need to be addressed.
First, as previously mentioned, an identical bound for the entanglement
obtains for the Klein-Gordon vacuum. The question arises as to whether
this reflects some sort of universality or is just an artifact of
our distillation protocol \cite{note2}. Naively, it might be expected
that the comparatively {}``poorer'' structure of the Dirac Hilbert
space, resulting from the anti-commutative nature of the field, leads
to a faster decay. However, the fact that our distillation protocol
is perturbative means that in the Klein-Gordon field case, the Hilbert
space's full structure does not come into play. This may very well
be the reason for the identical bound, but to say more would be pure
speculation.

Second, it is natural to ask whether the correlations giving rise
to this entanglement can be attributed to a local hidden-variable
model. In the Klein-Gordon field case, we were able to show \cite{VBI}
the detector's final state exhibits {}``hidden'' nonlocal correlations
\cite{Popescu}, in the sense that after local filtering \cite{Gisin}
an EPR state can be distilled. However, the same does not true here,
due to the presence of $\epsilon_{i}^{3}$ terms in the reduced density
matrix, Eq. (\ref{DM}), not present in the Bosonic equivalent, which
prevent the distillation of an EPR state. This question therefore
remains open.

The third question is how the results obtained change in the massive
case. Obviously, the presence of a mass term adds another scale to
the problem. From the point of view of our distillation protocol,
the Dirac equation no longer decouples and it is hard to see how use
can be made of a superoscillating function to satisfy the resulting
inequalities (without which we do not know how to distill entanglement
at arbitrarily long distances). Nonetheless, the fact that the main
contribution to the entanglement arises from high frequencies (see
\cite{VBI}) suggests that for a comparatively small mass our results
should remain unchanged \cite{note3}.

\section*{Appendix: Details of calculations}

To clarify some of the physical content behind the condition for entanglement,
Eq. (\ref{3+1 condition}), we outline here the important steps in
its derivation.

As already noted, in the absence of a mass term, the Dirac equation
decouples into a pair of equations for quanta of a definite handedness.
However, it is only in the Weyl representation of the Gamma matrices
\begin{equation}
\gamma_{0}=\left(\begin{array}{cc}
\mathbf{0} & \mathbf{1}\\
\mathbf{1} & \mathbf{0}\end{array}\right),\qquad\gamma_{i}=\left(\begin{array}{cc}
\mathbf{0} & \sigma_{i}\\
\sigma_{i} & \mathbf{0}\end{array}\right),\label{A1}\end{equation}
that these equations reduce from four component equations to two.
Taking the spinors to be normalised to unity we then have \begin{equation}
u_{r}\left(\vec{p}\right)=v_{r}\left(\vec{p}\right)=\frac{1}{\sqrt{2p\left(p-p_{z}\right)}}\left(\begin{array}{c}
p_{x}-ip_{y}\\
p-p_{z}\end{array}\right),\qquad u_{l}\left(\vec{p}\right)=v_{l}\left(\vec{p}\right)=u_{r}\left(-\vec{p}\right).\label{A2}\end{equation}
Focusing on the right handed vacuum, in terms of the Fourier expansion
of the field the emission and exchange terms are given by \begin{equation}
\left\Vert E_{i}\right\Vert ^{2}=\int\frac{d^{3}pd^{3}q}{\left(2\pi\right)^{6}}v_{r}^{\dagger}\left(\vec{q}\right)u_{r}\left(\vec{p}\right)u_{r}^{\dagger}\left(\vec{p}\right)v_{r}\left(\vec{q}\right)\left|\tilde{\epsilon}_{i}\left(\Omega_{i}+p+q\right)\right|^{2},\label{A3}\end{equation}
\begin{equation}
\left\langle 0\mid X_{AB}\right\rangle =-\int\frac{d^{3}pd^{3}q}{\left(2\pi\right)^{6}}v_{r}^{\dagger}\left(\vec{p}\right)u_{r}\left(\vec{q}\right)u_{r}^{\dagger}\left(\vec{q}\right)v_{r}\left(\vec{p}\right)e^{i\left(\vec{p}+\vec{q}\right)\cdot\left(\vec{x}_{A}-\vec{x}_{B}\right)}\tilde{\epsilon}_{A}\left(\Omega_{A}+p+q\right)\tilde{\epsilon}_{B}\left(\Omega_{B}-p-q\right),\label{A4}\end{equation}
where we have already carried out the temporal integration. Plugging
in the expressions for the spinors we get \begin{equation}
\left\Vert E_{i}\right\Vert ^{2}=\int\frac{d^{3}pd^{3}q}{\left(2\pi\right)^{6}}\frac{1}{2}\left(1+\frac{\vec{p}\cdot\vec{q}}{pq}\right)\left|\tilde{\epsilon}_{i}\left(\Omega_{i}+p+q\right)\right|^{2},\label{A5}\end{equation}
\begin{equation}
\left\langle 0\mid X_{AB}\right\rangle =-\int\frac{d^{3}pd^{3}q}{\left(2\pi\right)^{6}}\frac{1}{2}\left(1+\frac{\vec{p}\cdot\vec{q}}{pq}\right)e^{i\left(\vec{p}+\vec{q}\right)\cdot\left(\vec{x}_{A}-\vec{x}_{B}\right)}\tilde{\epsilon}_{A}\left(\Omega_{A}+p+q\right)\tilde{\epsilon}_{B}\left(\Omega_{B}-p-q\right).\label{A6}\end{equation}
In spherical coordinates the integration over angles is straightforward.
\begin{equation}
\left\Vert E_{i}\right\Vert ^{2}=2\int_{0}^{\infty}\frac{dpdq}{\left(2\pi\right)^{4}}p^{2}q^{2}\left|\tilde{\epsilon}_{i}\left(\Omega_{i}+p+q\right)\right|^{2},\label{A7}\end{equation}
\begin{eqnarray}
\left\langle 0\mid X_{AB}\right\rangle  & = & -2\int_{0}^{\infty}\frac{dpdq}{\left(2\pi\right)^{4}}\frac{pq}{L^{2}}\left[Cos\left(\left(p+q\right)L\right)-\frac{p}{L}Cos\left(pL\right)Sin\left(qL\right)-\frac{q}{L}Sin\left(pL\right)Cos\left(qL\right)\right.\nonumber \\
 &  & \left.+\frac{pq}{L^{2}}Sin\left(pL\right)Sin\left(qL\right)\right]\tilde{\epsilon}_{B}\left(\Omega_{A}+p+q\right)\tilde{\epsilon}_{B}\left(\Omega_{B}-p-q\right),\label{A8}\end{eqnarray}
where $L:=\left|\vec{x}_{A}-\vec{x}_{B}\right|$. If we now switch
to the variables $\omega=p+q$ and $\upsilon=p-q$, then integrating
over $\upsilon$, Eq. (\ref{3+1 condition}) quickly follows.

We note that the $Cos\left(\omega L\right)$ term on the LHS of Eq.
(\ref{3+1 condition}) arises from the angular integration over both
the $\frac{1}{2}$ and $\frac{\vec{p}\cdot\vec{q}}{2pq}$ terms resulting
from the spinor products. Were the $\frac{\vec{p}\cdot\vec{q}}{2pq}$
term absent, we would not be able to distill entanglement at any distance
$L$. It is interesting that this implies that we cannot realise our
distillation protocol in the real scalar vacuum via a square coupling.

\subsection*{Acknowledgments}

We would like to thank J. Kupferman for useful discussions. This work
was supported by the European Commission under the Integrated Project
Qubit Applications (QAP) funded by the IST directorate as contract
number 015848.


\begin{thebibliography}{99}
\bibitem{DM components}The bottom right component of the density matrix is of the fourth
order in the $\epsilon_{i}$, but is nonetheless retained to highlight
the similarity to the Bosonic vacuum.
\bibitem{note2}Note that the same bound, $\sim exp\left[-\left(L/R\right)^{2}\right]$,
obtains in 1+1 dimensions for the both the Klein-Gordon and Dirac
field.
\bibitem{note3}This requirement can be shown to be equivalent to $mR\ll1.$
\end{thebibliography}
\end{document}